\documentclass[twocolumn,amsmath,amssymb,floatfix]{revtex4}
\usepackage{graphicx}
\usepackage{dcolumn}
\usepackage{bm}
\begin{document}

\title{A Potential Energy Landscape Study\\ of the Amorphous-Amorphous
Transformation in H$_2$O}

\author{Nicolas Giovambattista$^1$,
H. Eugene Stanley$^1$ and Francesco Sciortino$^2$}

\affiliation{$^1$Center for Polymer Studies and Department of Physics,
Boston University, Boston, MA 02215 USA \\
$^2$ Dipartimento di Fisica and INFM Udr and 
Center for Statistical Mechanics and Complexity,
Universita' di Roma ``La Sapienza'' \\
Piazzale Aldo Moro 2, I-00185, Roma, Italy}

\begin{abstract}
We study the potential energy landscape explored during a
compression-decompression cycle for the SPC/E (extended simple point
charge) model of water.  During the cycle, the system changes from low
density amorphous ice (LDA) to high density amorphous ice (HDA). After
the cycle, the system does not return to the same region of the landscape,
supporting the interesting possibility that more than one significantly different
configuration corresponds to LDA.  We find that the regions of the
landscape explored during this transition have properties remarkably
different from those explored in thermal equilibrium in the liquid
phase. 

\end{abstract}

\pacs{PACS numbers: XXX} 

\date{Received 21 Feb 2003, Revised 5 June 2003. --- LB9394 ---}

\maketitle

The physics of water in its supercooled and glassy phases is the object
of several recent experimental 
\cite{mishjcp,mishimaLL,loerting2,science,finney,loerting} and
theoretical \cite{tse,poole,poolepre,tseNature}
investigations. One of the most fascinating aspects is the possibility
of a liquid-liquid phase transition \cite{poole} and the relation
between the two liquid phases and the glassy phases of water
 \cite{poolepre}. These glassy phases are produced via routes as
different as crystal compression, hyper-quenching, and vapor deposition,
and are characterized by different densities and by different local
structures \cite{mishimaNature85,loerting,science}. The intrinsic
out-of-equilibrium nature of these phases leads to some ambiguities in
the identification of the materials produced via different routes,
and recent debate concerns the classification of the
different amorphous structures found \cite{loerting,loerting2,finney,science}.
High density amorphous ice (HDA) resulting from the compression of
hexagonal ice at 77 K can be converted to low density amorphous ice
(LDA) by releasing the pressure at $\approx115$~K
 \cite{mishima84,handa,floriano}. HDA and LDA can be inter-converted via an
 apparent first-order transition by applying or releasing pressure
 \cite{mishimaNature85,mishimaVisual}. Molecular dynamics simulations, based 
on different model potentials\cite{models},  have been able to reproduce the 
qualitative features of the crystalline ice-HDA, HDA-LDA, and LDA-HDA
transitions \cite{tse,poolepre}, even if the compression rates in simulations
 are several orders of magnitude larger than in experiments.

To study the LDA-HDA transition and the relation between glassy and liquid
  water we use tools that have been recently developed to study
 out-of-equilibrium  liquids \cite{mossaaging} in the potential energy
 landscape (PEL) framework \cite{Goldstein,stillweber,debenedetti}.
 In the PEL approach,  the $6N$-dimensional configurational space --- 
defined by the $3N$ center of mass coordinates and by the $3N$ Euler 
angles, where $N$ is the number of molecules  --- is  partitioned into 
a set of basins, each of them associated with a different local minimum of 
the potential energy landscape. The set of points belonging to the same basin
 are those which, under a steepest descent minimization procedure, end up
  in the same local minimum.  The local minima in the PEL are called inherent
 structures (IS).  Each IS configuration is characterized by  its potential
 energy ($e_{IS}$), its pressure ($P_{\mbox{\scriptsize IS}}$) and by the 
$6N$ local curvatures of the PEL, which can be estimated in the harmonic 
approximation by diagonalizing the Hessian matrix \cite{footnote}. Basins
 with different depths have different curvatures
\cite{kstepl,sastrynature,mossaotp,fstarr}. 
One possible  measure  of the average basin curvature 
--- the one used in the present work ---  is offered by  the ``shape function''
\begin{equation}
{\cal S}_{\mbox{\scriptsize IS}} \equiv \frac{1}{N} \sum_{i=1}^{6N-3}
\ln\left(\frac{\hbar\omega_i}{A_0}\right),
\end{equation}
where $\omega_i$ is the frequency of vibrational mode $i$ \cite{footnote},
 $\hbar$ is the Planck constant, and $A_0=1$~kJ/mol. This choice is motivated
 by
 the fact that ${\cal S}_{\mbox{\scriptsize IS}} $ enters into the evaluation
 of the basin free energy in the harmonic approximation \cite{st}.

Recent work has provided a detailed statistical description of the number, depth  
and shape of the basins, for models of both simple liquids 
\cite{st,sastrynature,ivan} and molecular liquids 
\cite{mossaotp,francisspce}. It has
been shown that, on cooling, the liquid samples regions of the PEL 
characterized by lower and lower $e_{IS}$ values \cite{debenedetti} 
and that both $P_{\mbox{\scriptsize IS}}$ 
and ${\cal S}_{\mbox{\scriptsize IS}}$ 
are correlated with $e_{\mbox{\scriptsize IS}}$ \cite{eos}.
Numerical studies of the PEL sampled by the 
equilibrium liquid  permit  precise calculations of both 
${\cal S}_{\mbox{\scriptsize IS}}(e_{\mbox{\scriptsize IS}},\rho)$ and
$P_{\mbox{\scriptsize IS}}(e_{\mbox{\scriptsize IS}},\rho)$, providing a
detailed description of the region of the PEL sampled under equilibrium
conditions \cite{nota,fstarr,pablojcp}. 

In this Letter we aim at comparing the properties of the PEL sampled during
the LDA-HDA transition with the properties of the PEL explored by the
 equilibrium liquid.  If the regions of PEL explored by the glass during
 the transition are identical \cite{identical}  to the regions explored by
 the equilibrium liquid, we are entitled to connect the glass-glass 
transformation (GGT) to a transition taking place in the liquid state,
 supporting the hypothesis that the LDA-HDA transformation is an 
out-of-equilibrium manifestation of a liquid-liquid first order transition.
  We discover that during the transformation the
 system explores regions of the landscape which are never
 explored in equilibrium, supporting  
the possibility that the LDA-HDA transformation  and the  liquid-liquid
 first order transition might be independent phenomena.  In this respect,
 the LDA-HDA transformation observed in numerical simulations 
\cite{poole,tse,poolepre} should not
 be taken as proof of the existence of a liquid-liquid transition.

We perform molecular dynamics (MD) simulations of 
216 molecules interacting via the  simple point charge
extended (SPC/E) model of water \cite{berendsen}. This model has been
studied extensively; the $\rho$ and $T$ dependence of structural and
dynamic properties in equilibrium have been calculated. 
We use a  simulation time step of 1fs and long range forces are handled using
 the reaction field method. We identify 
the IS by minimizing the potential energy using a conjugate gradient 
minimization algorithm and calculate  $e_{\mbox{\scriptsize IS}}$, 
${\cal S}_{\mbox{\scriptsize IS}}$, and $P_{\mbox{\scriptsize IS}}$ 
 in the resulting local minimum configuration.
During the MD simulation, starting equilibrium configurations at 
$\rho=0.9$ g/cm$^3$ are either (a) decompressed to 0.8 g/cm$^3$ or (b)
compressed to 1.4 g/cm$^3$ and then decompressed to 0.8 g/cm$^3$. We simulate at both $T=0$~K and $T=77$~K. 

At $77$~K we perform MD simulations using two different
compression/decompression rates ($d\rho/dt=5 \times 10^{-4}$g/cm$^3$/ps and $5\times 10^{-5}$g/cm$^3$/ps) and average over 16 different
realizations.
In the calculation at $T=0$~K, each step consists of a density change of either $\Delta\rho=5\times 10^{-4}$g/cm$^3$ or $\Delta\rho=5\times 10^{-5}$g/cm$^3$, followed by an energy minimization. At each step the system is compressed by $\Delta\rho$ and the center of mass of each molecule is isotropically scaled. 
Initial configurations are extracted from a pre-existing ensemble of IS
generated by quenching equilibrium liquid configurations at $\rho=0.90$
g/cm$^3$ and $T=220$ K. At this state point, SPC/E  describes the LDA
structure accurately.   Any low-temperature equilibrium configuration with
density close to the LDA density could be used as a starting configuration.

\begin{figure}[t]
\centering
\includegraphics[width=3.5in]{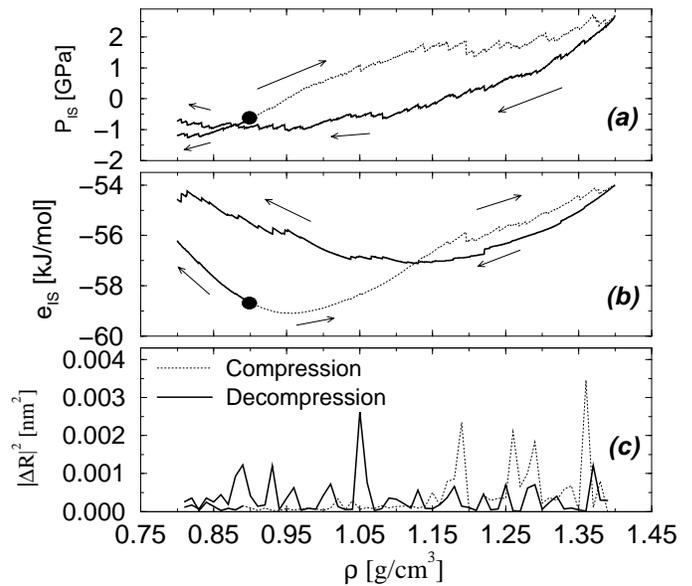}
\caption{(a) $P_{\mbox{\scriptsize IS}}$ and (b) 
$e_{\mbox{\scriptsize IS}}$ for $T=0$~K as functions of $\rho$, where a starting glass
configuration (black dot) is decompressed to $\rho=0.8$, and also
compressed to $\rho=1.4$~g/cm$^3$ and then decompressed to
$\rho=0.8$g/cm$^3$. The density in each step of the
 compression/decompression cycle is changed 
by $\Delta\rho=5\times 10^{-5}$g/cm$^3$. (c) Squared distance 
between a sequence of IS configurations differing by $10^{-2}$g/cm$^3$ along the
compression/decompression path. Note the correlation between the large
molecular rearrangements in part (c) and the corresponding
discontinuities in $P_{\mbox{\scriptsize IS}}$ and 
$e_{\mbox{\scriptsize IS}}$ in part (a) and (b), such as one finds at $\rho=1.20$g/cm$^3$.}
\label{fig:1}
\end{figure}

Figure~\ref{fig:1} shows the behavior of $P_{\mbox{\scriptsize IS}}$ and
$e_{\mbox{\scriptsize IS}}$ for the compression and decompression of a
single $T=0$~K configuration. Following each density change, the system
is displaced from the local minimum in the PEL and a steepest descent
minimization is performed to bring the system to the new location of the
minimum. $e_{\mbox{\scriptsize IS}}$ and $P_{\mbox{\scriptsize IS}}$ are continuously modified under the density changes, and the initial parts of the curves
in Figs.~\ref{fig:1}a and \ref{fig:1}b show such continuous
modification. Above $\rho\approx 1.05$~g/cm$^3$, this continuous, smooth process is interrupted by sudden changes in both $P_{\mbox{\scriptsize
IS}}$ and $e_{\mbox{\scriptsize IS}}$, clearly visible in
Fig.~\ref{fig:1}. Figure~\ref{fig:1}c shows the squared distance between
two IS differing by $10^{-2}$~g/cm$^3$, 
%
$|\Delta R|^2 \equiv \frac{1}{N} \sum_{i=1}^Nd\mathbf{r}_i^2$,
%
where $d\mathbf{r}_i$ is the displacement of the center of mass of the \textit{i}-th
molecule. We see that $|\Delta R|^2$ changes significantly when 
discontinuous jumps in $P_{\mbox{\scriptsize IS}}$ and
$e_{\mbox{\scriptsize IS}}$ occur. Thus, as in silica \cite{lacks,lacksSilica},
these changes are due to a mechanical instability associated with the
vanishing of the lowest frequency mode, which forces the system to
abandon the previous unstable configuration in favor of a new basin.

\begin{figure}[t]
\centering
\centerline{\includegraphics[width=3.7in]{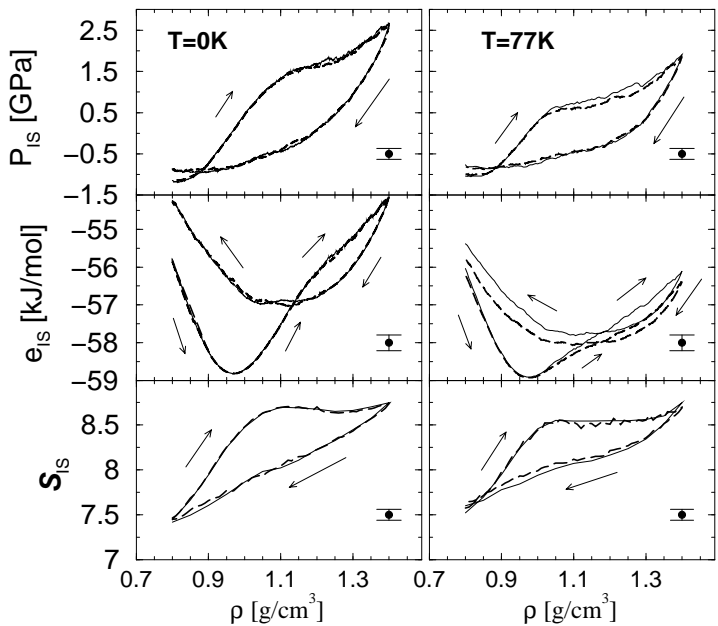}}
\caption{ $P_{\mbox{\scriptsize IS}}$, $e_{\mbox{\scriptsize IS}}$, and
${\cal S}_{\mbox{\scriptsize IS}}$, averaged over 16 different
realizations, as functions of $\rho$ during the compression and
decompression at $T=0$~K (left panels) and $T=77$~K (right panels) for
the fast (solid line) and slow (dashed line) compression rates. Error bars
are shown on the left lower corner.}
\label{fig:2}
\end{figure}

Figure~\ref{fig:2} shows $P_{\mbox{\scriptsize IS}}$,
$e_{\mbox{\scriptsize IS}}$, and ${\cal S}_{\mbox{\scriptsize IS}}$ for
both $T=0$~K and $T=77$~K, 
averaged over 16 different realizations as functions of $\rho$ during
compression and decompression at two different rates. The
$P_{\mbox{\scriptsize IS}}$ curve shows the typical behavior found for
the pressure $P$ observed in previous studies of compression and
decompression of tetrahedral glasses
\cite{tse,poole,floriano,mishjcp}. We find a strong hysteresis, an
observation which has been interpreted as evidence in favor of a first
order transition between two distinct structures, LDA and HDA
\cite{mishjcp,tse,poole}.

On compressing, a significant flattening of $P_{\mbox{\scriptsize IS}}$
is observed when the density becomes larger than $\approx 1.1$~g/cm$^3$
at $T=0$~K and larger than $\approx 1.0$~g/cm$^3$ at $T=77$~K. At the
beginning of the compression, the compression rate does not appear to
play a significant role, but when the system is forced to change basins
(due to thermal effects or mechanical instabilities) the compression
rate becomes quite relevant. For the smaller compression rate,
$P_{\mbox{\scriptsize IS}}$ and $e_{\mbox{\scriptsize IS}}$ are slightly
smaller. 

We next compare our results for $T=0$~K and $T=77$~K. Transitions at
$T=0$~K are driven by the vanishing of the lowest normal mode
frequencies resulting in the compression-induced disappearance of the
explored basin. From the top panel of Fig.~\ref{fig:2} we see that even
though both systems start from the same configuration (and hence same
IS) at $\rho=0.9$~g/cm$^3$, the additional thermal energy of the 77 K
system enables the escape from the starting basin to occur at a smaller
density. This results in a more effective relaxation process and,
correspondingly, in a flattening of $P_{\mbox{\scriptsize IS}}$ at a
smaller density value (compared to the $T=0$~K case) and in smaller
$P_{\mbox{\scriptsize IS}}$ and $e_{\mbox{\scriptsize IS}}$ values. The
shape of the basin is also significantly modified by the
compression/decompression cycle. We also find that the basin shape 
${\cal S}_{\mbox{\scriptsize IS}}$ increases on compression and decreases on
decompression, coherent with the frequency shifts of the translational
peaks of the density of states (i.e. in the probability distribution for $\omega_i$). In the region where
$P_{\mbox{\scriptsize IS}}$ flattens, neither 
${\cal S}_{\mbox{\scriptsize IS}}$ nor the density of states changes
significantly.

After a complete compression/decompression cycle, the system does not
return to the same configuration. Indeed, Fig.~\ref{fig:2} shows that
 while $P_{\mbox{\scriptsize IS}}$ has the same value on compression and
decompression at $\rho \approx 0.89$~g/cm$^3$, the values for
$e_{\mbox{\scriptsize IS}}$ and ${\cal S}_{\mbox{\scriptsize IS}}$
differ---i.e., at $\rho \approx 0.89$~g/cm$^3$, $T$ and $P$ are
identical, but quite significant differences are detected at a
microscopic level. This finding supports the interesting possibility
that more than one significantly different configuration corresponds to
LDA.

\begin{figure}[t]
\centering
\centerline {\includegraphics[width=3.2in]{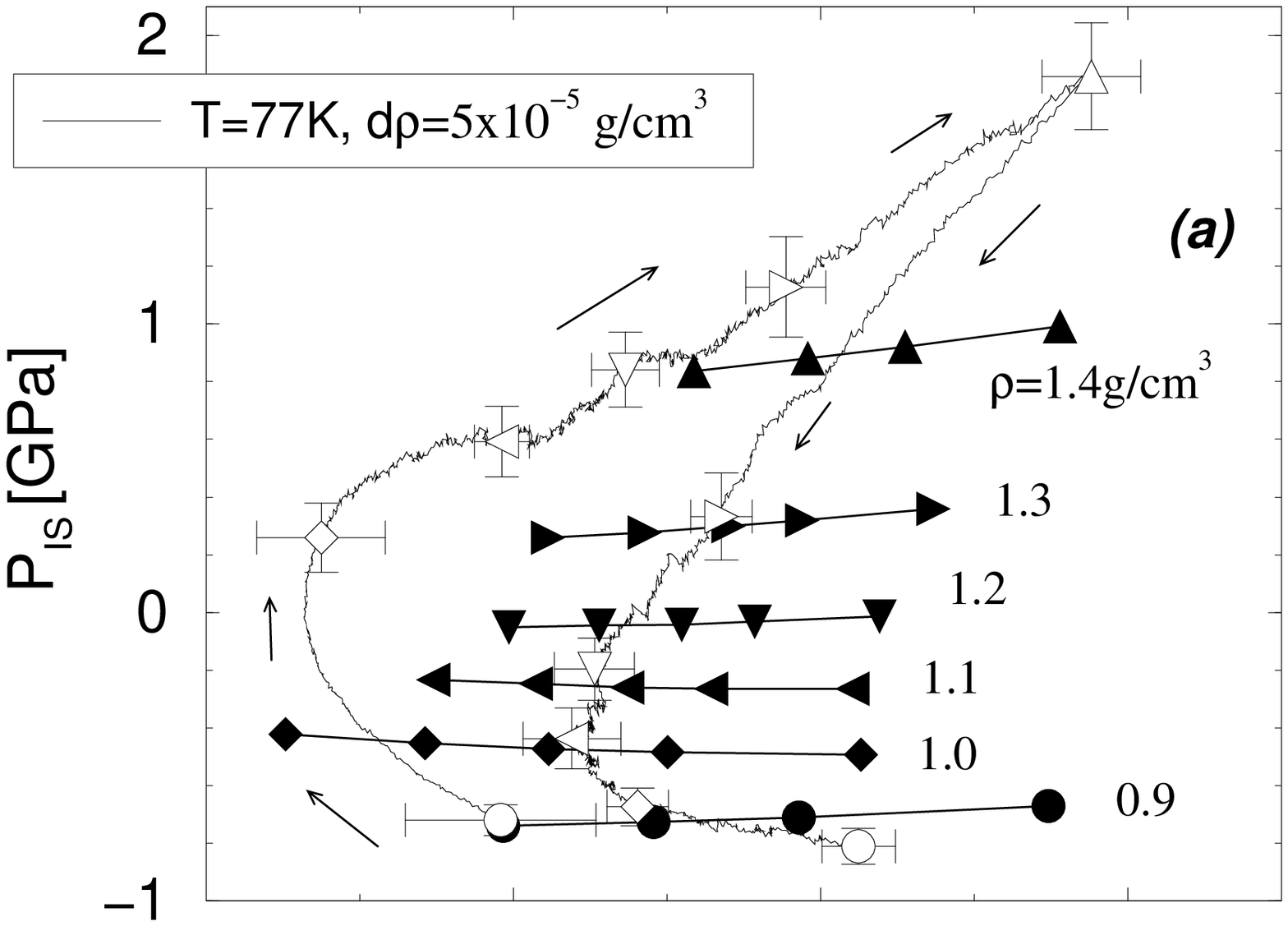}}
\centerline{
\includegraphics[width=3.2in]{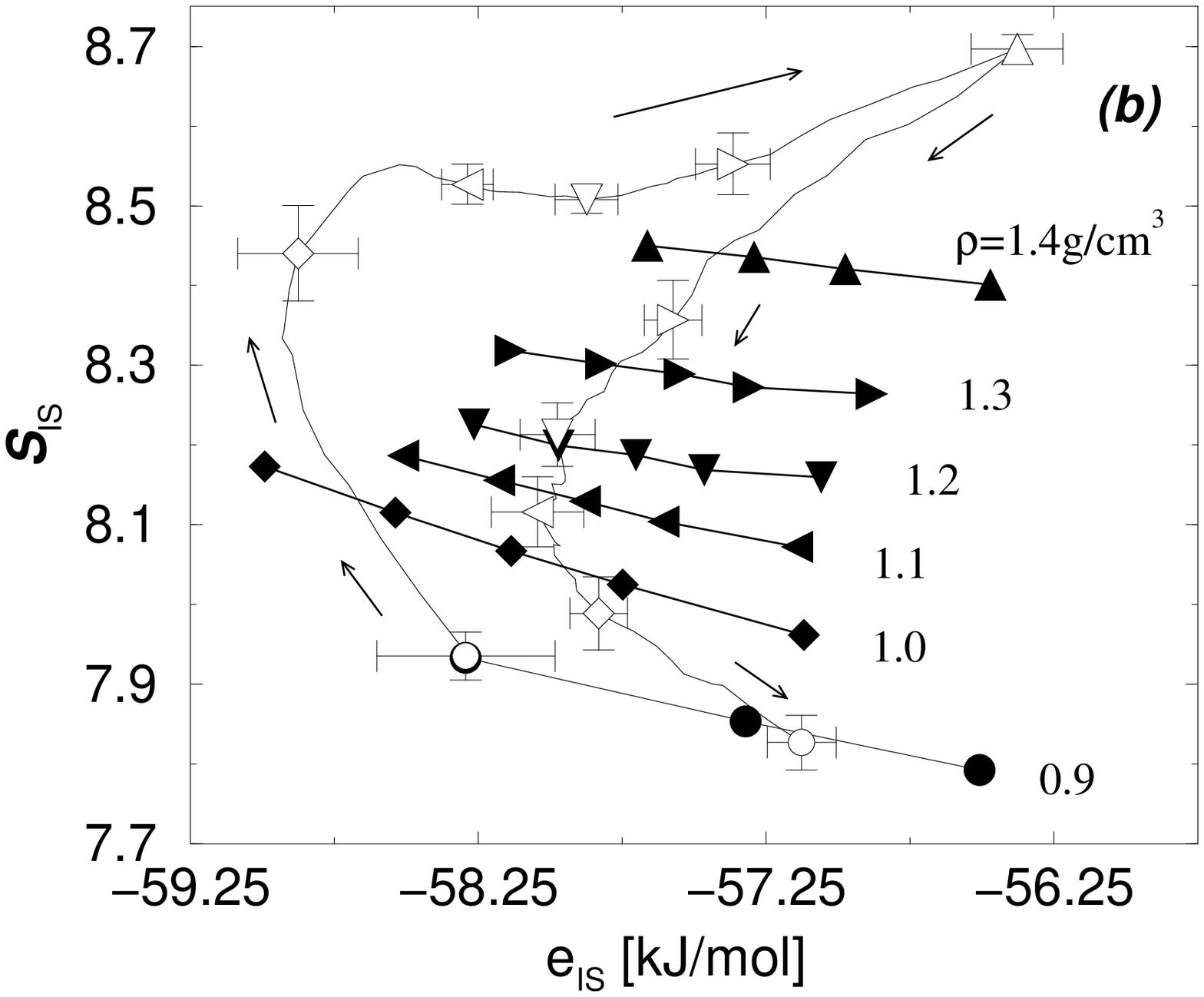}
}
\caption{(a) $P_{\mbox{\scriptsize IS}}$ and (b) 
${\cal S}_{\mbox{\scriptsize IS}}$ as functions of $e_{\mbox{\scriptsize IS}}$
at different densities for the equilibrium liquid (filled symbols) and
for the compression/decompression of glass (thin line) at both
temperatures studied. Open symbols indicate the density during the
compression/decompression. The relative location of the open symbol with
respect to the location of the same filled symbol is a measure of the
difference between the landscape properties in equilibrium and in the
glass. The significant difference in the location confirms that the
compressed/decompressed glass explores PEL regions not explored under
equilibrium conditions at the same density. Densities are $0.9(\bigcirc)$,
 $1.0 (\Diamond)$ , $1.1 (\triangleleft)$, $1.2(\bigtriangledown)$,
 $1.3 (\rhd)$, and $1.4(\bigtriangleup)$~g/cm$^3$.
 The error bars for these densities are indicated in the figures. Data for
equilibrium liquid are averaged over 2000 realizations and error bars are
smaller than the corresponding symbol size.}
\label{fig:3}
\end{figure}

We next compare PEL properties sampled in equilibrium at different $T$
and $\rho$ with PEL properties sampled by the glass during the
compression/decompression path with the slow compression rate. Figure~3
contrasts $P_{\mbox{\scriptsize IS}}$ and 
${\cal S}_{\mbox{\scriptsize IS}}$ in an equilibrium liquid (solid symbols)
\cite{nota} and during the
compression/decompression runs (open symbols) at $T=77$ K. Surprisingly,
we find that even a slight density change (from $\rho=0.9$ to $1.0$
g/cm$^3$) allows the system to begin exploring regions of configuration
space not explored under equilibrium conditions at the corresponding
density. For a given $\rho$, $P_{\mbox{\scriptsize IS}}$ and 
${\cal S}_{\mbox{\scriptsize IS}}$ are larger in the compressed glass than
observed under equilibrium liquid conditions. We find that the $T=0$ K
results show even larger differences.

For small density changes ($\rho \lesssim1.1$~g/cm$^3$ at $T=0$~K, or
for $\rho \lesssim1.0$~g/cm$^3$ at $T=77$~K) the system is still
confined in the (deformed) starting basin. In this density range (i.e.,
before any basin change) compression rate effects do not play any role.
Figure~\ref{fig:3} shows that even before basin change effects become
relevant, the system is already in a compressed state which hardly
remembers an equilibrium liquid configuration. The values of
$P_{\mbox{\scriptsize IS}}$ and ${\cal S}_{\mbox{\scriptsize IS}}$ at
$\rho=1.0$~g/cm$^3$ are comparable to the equilibrium liquid values
corresponding to a much higher density of 1.3~g/cm$^3$. Hence we
conclude that the flattening of the $P_{\mbox{\scriptsize IS}}$ curve of
Fig.~\ref{fig:2}---associated with the LDA-HDA transition---takes place
in a region of the PEL that is never explored under equilibrium
conditions.

Our results support the view that the compression/decompression of
glassy water takes place along a PEL path rarely explored by the
equilibrium liquid. Such information ---  which
  could not have been extracted from standard analysis
 (such as standard thermodynamic or structural quantities)
 ---  is made possible by the extreme sensitivity of the landscape
 properties.  We have shown that the IS visited by the liquid (on the
 time scale probed in computer simulations) are different than those
 sampled by the glass. This observation has profound
 consequences for the possibility of developing a thermodynamic approach
 to the LDA-HDA transformation. Indeed, only when the glass state can be
 associated with a frozen liquid (i.e. when the equilibrium relations 
${\cal S}_{\mbox{\scriptsize IS}}(e_{\mbox{\scriptsize IS}},\rho)$ and
$P_{\mbox{\scriptsize IS}}(e_{\mbox{\scriptsize IS}},\rho)$ are satisfied)
 it becomes possible to provide a thermodynamic description of the glass and
 formally connect the  mechanical instability of the LDA-HDA transformation 
to a mechanical instability in the liquid state\cite{st,mossaaging}.
If  the liquid  and the glass had shared the same portion of the
PEL, then the mechanical instability observed during the transformation
 would have constituted genuine evidence of a  liquid-liquid transition.  
The observation that the  LDA-HDA transformation is unrelated to properties
 of the equilibrium liquid \cite{jagla}  does not allow us to relate the
  observed numerical GGT  to a liquid-liquid transition. Due to the extremely 
different time scales probed in experiments and in simulations,
our numerical results do not exclude neither the existence  of a liquid-liquid 
transition nor the possibility that the LDA-HDA transition observed
experimentally
 is related to a liquid-liquid transition. 
Our results show
 that the PEL properties at the beginning and the end of the
 compression/decompression cycle are  different, suggesting that 
 several distinct configurations can be associated with LDA.

\subsubsection*{Acknowledgments}

We thank NSF Chemistry Program, MIUR Cofin 2002 and Firb and INFM Pra GenFdt for support
and SHARCNET and BU Computation Center for a generous allocation of CPU
time. FS thanks UWO for its hospitality.


\begin{references}

\bibitem{mishjcp} O. Mishima, J. Chem. Phys. {\bf 100}, 5910 (1993). 

\bibitem{mishimaLL} O. Mishima and H. E. Stanley, Nature {\bf 392}, 164
(1998).

\bibitem{loerting2} T. Loerting {\it et al.}, 
Phys. Chem. Chem. Phys. {\bf 3}, 5355 (2001). 

\bibitem{science} C. A. Tulk {\it et al.}, 
Science {\bf 297}, 1320 (2002).

\bibitem{loerting} T. Loerting {\it et al.}, 
J. Chem. Phys. {\bf 116}, 3171 (2002).

\bibitem{finney} J. L. Finney {\it et al.}, 
Phys. Rev. Lett. {\bf 89}, 205503 (2002).





\bibitem{tse} J. S. Tse and M. L. Klein, Phys. Rev. Lett. {\bf 58}, 1672
(1987). 

\bibitem{poole} P. H. Poole {\it et al.}, 
Nature {\bf 360}, 324 (1992). 

\bibitem{poolepre} P. H. Poole {\it et al.}, 
Phys. Rev. E {\bf 48}, 4605 (1993). 

\bibitem{tseNature} J. S. Tse {\it et al.}, Nature {\bf 400}, 647 (1999). 

\bibitem{mishimaNature85} O. Mishima {\it et al.}, 
Nature {\bf 314}, 76 (1985). 

\bibitem{mishima84}
O. Mishima {\it et al.},  
Nature {\bf 310}, 393 (1984).

\bibitem{handa} Y. P. Handa, {\it et al.}, 
J. Chem. Phys.  {\bf 84}, 2766 (1986). 

\bibitem{floriano} M. A. Floriano {\it et al.}, 
J. Chem. Phys. {\bf 91}, 7187 (1989). 

\bibitem{mishimaVisual} O. Mishima {\it et al.},   
Science {\bf 254}, 406 (1991). 


\bibitem{models}
Previous simulations on the LDA-HDA
transformation were done using the TIP4P\protect\cite{tse}  and ST2\cite{poolepre}  models.


\bibitem{mossaaging} S. Mossa {\it et al.}, 
Eur. Phys. J. B {\bf 30}, 351 (2002).


\bibitem{Goldstein} M. Goldstein, J. Chem. Phys. {\bf 51}, 3728 (1969). 

\bibitem{stillweber} F. H. Stillinger and T. A. Weber, Phys. Rev. A {\bf
28}, 2408 (1983); C. A. Angell, Science {\bf 267}, 1924 (1995). 


\bibitem{debenedetti} S. Sastry {\it et al.},
Nature {\bf 393}, 554 (1998). 











\bibitem{footnote}
To obtain $\omega_i$, we first  calculate the $6N \times 6N$ matrix of  the second derivatives of the potential energy with respect to the $6N$ coordinates of the molecules, evaluated at the IS configuration. Then we diagonalize the matrix. The $i$-th eigenvalue is  $\omega^2_i$.
 Physically it indicates the local curvature of the PEL at the IS along the
 direction of the $i$-th eigenvector of the matrix. The
 probability distribution for $\omega_i$ defines the vibrational density of states.


\bibitem{kstepl} W. Kob {\it et al.}, Eur. Phys. Lett. {\bf 49}, 590 (2000).

\bibitem{fstarr}
F.W. Starr {\it et al.}, Phys. Rev. Lett. {\bf 63}, 041201 (2001).

\bibitem{sastrynature} S. Sastry, Nature {\bf409}, 164 (2001). 

\bibitem{mossaotp} S. Mossa {\it et al.}, 
Phys. Rev. E {\bf 65}, 041205 (2002). 

\bibitem{st} F. Sciortino and P. Tartaglia, Phys. Rev. Lett. {\bf 86},
 107 (2001); F. Sciortino {\it et al.},
{\it ibid.} {\bf 83}, 3214 (1999). 

\bibitem{ivan} I. Saika-Voivod {\it et al.}, Nature {\bf 412}, 514 (2001).

\bibitem{francisspce} F. W. Starr {\it et al.}, 
Phys. Rev. E {\bf 63}, 041201 (2001). 

\bibitem{eos} E. La Nave {\it et al.}, 
Phys. Rev. Lett. {\bf 88}, 225701 (2002). 

\bibitem{nota}
E. La Nave {\it et al.}, preprint. The relations  
${\cal S}_{\mbox{\scriptsize IS}}(e_{\mbox{\scriptsize IS}},\rho)$ and
$P_{\mbox{\scriptsize IS}}(e_{\mbox{\scriptsize IS}},\rho)$ from
equilibrium systems are estimated along an isochore as follows. First, we
generate set of 2000 independent equilibrium configuration, with standard
MD, at several different $T$. All 
configurations are then minimized to generate the associated IS. 
  For each $T$, the average values of $P_{\mbox{\scriptsize IS}}$,
 ${\cal S}_{\mbox{\scriptsize IS}}$  and  $e_{\mbox{\scriptsize IS}}$
are then calculated, averaging over the generated inherent structures set.

\bibitem{pablojcp} 
P.G. Debenedetti {\it et al.}, J. Phys. Chem. B {\bf 103}, 7390 (1999).



\bibitem{identical} 
The word  {\it identical} means that $e_{\mbox{\scriptsize IS}}$, 
${\cal S}_{\mbox{\scriptsize IS}}$, and $P_{\mbox{\scriptsize IS}}$ along
 the GGT satisfy  the equilibrium functional forms 
${\cal S}_{\mbox{\scriptsize IS}}(e_{\mbox{\scriptsize IS}},\rho)$ and
$P_{\mbox{\scriptsize IS}}(e_{\mbox{\scriptsize IS}},\rho)$.

\bibitem{berendsen} H. J. Berendsen {\it et al.}, 
J. Phys. Chem. {\bf 91}, 6269 (1987). 


\bibitem{lacks} D. L. Malandro and D. J. Lacks, J. Chem. Phys. {\bf107},
5804 (1997); 
D. J. Lacks, Phys. Rev. Lett. {\bf 80}, 5385 (1998); 


\bibitem{lacksSilica} D. J. Lacks, Phys. Rev. Lett. {\bf 84}, 4629
(2000). 



\bibitem{jagla} E. A. Jagla, Phys. Rev. Lett. {\bf 86}, 3206 (2001).





\end{references}
\end{document}